\documentclass[a4paper]{jpconf}
\usepackage{amsmath}
\usepackage{amssymb}
\usepackage{bm}
\usepackage{graphicx}
\begin{document}
\title{Quantum Magnets and Matrix Lorenz Systems}

\author{J Tranchida$^{1,2}$, P Thibaudeau$^1$ and S Nicolis$^2$}

\address{$^1$ CEA DAM/Le Ripault, BP 16, F-37260, Monts, France}
\address{$^2$ CNRS--Laboratore de Math\'ematiques et Physique Th\'dorique (UMR7350)\\
F\'ed\'eration de Recherche ``Denis Poisson'' (FR2964)\\
Universit\'e ``Fran\c{c}ois Rabelais'' de Tours\\
Parc Grandmont, 37200 Tours, France} 

\ead{julien.tranchida@cea.fr,pascal.thibaudeau@cea.fr,stam.nicolis@lmpt.univ-tours.fr}

\begin{abstract}
The Landau--Lifshitz--Gilbert equations for the evolution of the magnetization, in presence of an external torque,  can be cast in the form of the Lorenz equations and, thus, can describe chaotic fluctuations.
To study quantum effects, we describe the magnetization by matrices, that take values in a Lie algebra. The finite dimensionality of the representation encodes the quantum fluctuations, while the non-linear nature of the equations can describe chaotic fluctuations. We identify a criterion, for the appearance of such non-linear terms. This depends on whether an invariant, symmetric tensor of the algebra can vanish or not. This proposal is studied in detail for the fundamental representation of  $\mathfrak{u}(2)=\mathfrak{u}(1)\times\mathfrak{su}(2)$. We find a knotted structure for  the attractor, a bimodal distribution for the largest Lyapunov exponent and that the dynamics takes place within the Cartan subalgebra, that does not contain only the identity matrix, thereby can describe the quantum fluctuations.  
\end{abstract}

\section{Introduction}
Recent advances in magnetic materials and techniques allow manipulation of spin moments at nanoscale resolution. Thus, chaotic fluctuations become 
significant and their control have been the subject of both experimental and theoretical extensive studies \cite{Wigen:1994hc}. One of the first direct observation of period doubling and chaos was spin-wave instabilities in yttrium iron garnet (YIG) and has been documented more than thirty years ago \cite{Gibson:1984fu}. Using the ferromagnetic resonance technique, several routes to chaos have been found and explored including periodic-doubling cascades, quasi periodic and intermittent dynamics which exhibit complex magnetic behaviors \cite{Fernandez-Alvarez:2000ij,Mayergoyz:2009kl}. Recently, the phase diagram of a chaotic magnetic nanoparticle has been presented \cite{Bragard:2011nx}, which was obtained by monitoring the classical dynamics of its magnetization, modeled by the Landau-Lifshitz-Gilbert (LLG) equation.

What has received much less attention is the contribution of {\em quantum} fluctuations, that become significant at nanoscale resolution and are crucial for controlling qubit devices~\cite{boissoneault}. These might affect non-linear effects in new ways.

The challenge, therefore, is to describe the interplay between the two sources of fluctuations in a way that can lead to a deeper understanding of their effects and predict new features. To this end a model is proposed that displays both and allows us to distinguish them in a particularly clean fashion.

\begin{figure}[thp]
\begin{center}\resizebox{0.4\columnwidth}{!}{\includegraphics{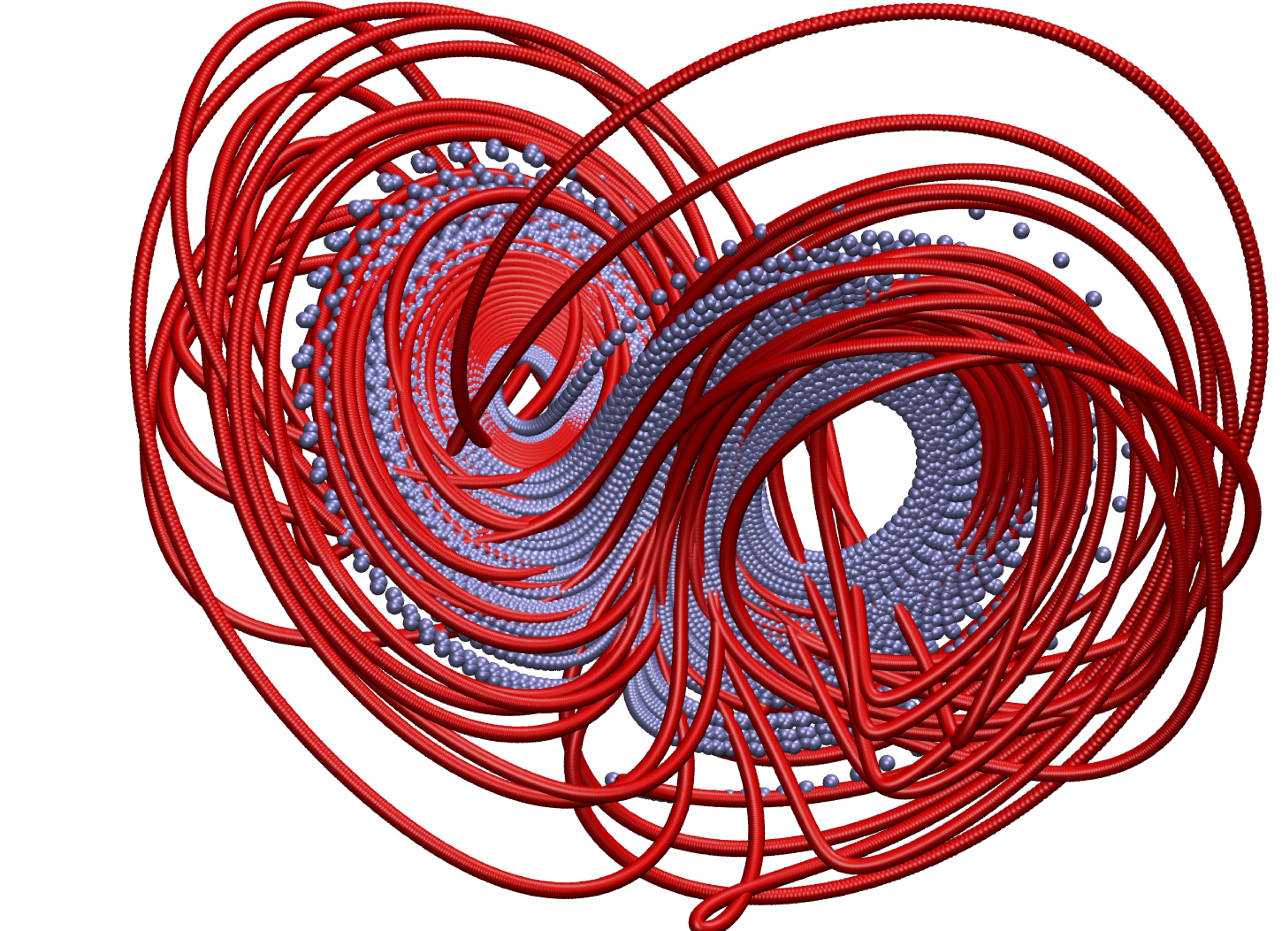}}\end{center}
\caption{(Color online) Plot of the parametric solution of the time evolution of $\mathrm{Tr}(X), \mathrm{Tr}(Y)\, \mathrm{and}\, \mathrm{Tr}(Z)$ 
with $\sigma=10$, $r=28$, $b=8/3$. Classical Lorenz system in gray and its matrix version in red. \label{Fig1}}
\end{figure}

Let take a closer look to the LLG equation. The motion of the magnetization ${\vec M}$ in 3D space that spins around an effective pulsation ${\vec\omega}$ with a damping vector ${\vec D}$ and an external torque $\vec{T}$, is given by
\begin{equation}
\dot{\vec M}={\vec\omega}\times{\vec M}-{\vec D}+\vec{T}\label{LLG}.
\end{equation} 
This equation can be used to describe a particular magnetic system with 3 single-ion anisotropy axes $\vec{n}_i$ of intensity $\eta_i$, under a static magnetic field $\vec{B}$ and a diagonal Bloch-Blombergen damping $\vec{D}=(M_1/\tau_1,M_2/\tau_2,M_3/\tau_3)$ with the torque $\vec{T}=(0,0,d)$ along the z-axis. In such a situation, $\vec{\omega}=\sum_{i=1}^3\eta_i(\vec{n}_i.\vec{M})\vec{n}_i-\vec{\beta}$ with $\eta_i=2K_i/\mu_0M_s^2$, $\vec{\beta}=\vec{B}/\mu_0M_s$, $K_i$ is the anisotropy energy along the $i$-axis and $M_s$ is the saturation magnetization. 
Equation (\ref{LLG}) can be written in components as 
\begin{equation}
\begin{array}{rcl}
\dot{M}_1&=&\displaystyle{(\eta_2-\eta_3)M_2M_3-\beta_2M_3+\beta_3M_2-\frac{M_1}{\tau_1}}\\
\dot{M}_2&=&\displaystyle{(\eta_3-\eta_1)M_3M_1-\beta_3M_1+\beta_1M_3-\frac{M_2}{\tau_2}}\\
\dot{M}_3&=&\displaystyle{(\eta_1-\eta_2)M_1M_2-\beta_1M_2+\beta_2M_1-\frac{M_3}{\tau_3}+d}
\end{array}
\label{LLG2}
\end{equation}
The linear transformation 
\begin{equation}
\label{LLG-lor}
\begin{array}{l}
\displaystyle
M_1 = x\\
\displaystyle
M_2 = y\\
\displaystyle
M_3 = z-r-\sigma\\
\end{array}
\end{equation}
when $\eta_1=2,\eta_2=1,\eta_3=1,\beta_1=0,\beta_2=0,\beta_3=\sigma,\tau_1=1/\sigma,\tau_2=1,\tau_3=1/b, d=-b(r+\sigma)$, lets the Landau--Lifshitz--Gilbert equations~(\ref{LLG2}) take the form of  the well known prototypical system that displays the full repertoire of behaviors from regular to chaotic, along many routes, namely the Lorenz system~\cite{Lorenz:1963vn}
\begin{equation}
\begin{array}{ccl}
\dot{x}&=&\sigma(y-x)\\
\dot{y}&=&x(r-z)-y\\
\dot{z}&=&xy-bz
\end{array}
\label{CLS}
\end{equation}
whose typical solution (in the chaotic phase) is displayed in Fig.~\ref{Fig1}. 
Here $\dot{x}\equiv dx/dt$ and similarly for the other variables. While these equations were deduced to describe classical fluid dynamics and the original parameters reflect this fact: $\sigma$ is the Prandtl number, $r$ the Rayleigh number and $b$ is the aspect ratio of the ``cell'' (real or virtual), their scope is, in fact, much broader, as was realized from the work in the 70's~\cite{Feigenbaum:1978fk}. These are, still, classical equations and the variables $(x,y,z)$ are commuting quantities. 

To describe the quantum fluctuations, a generalization of these equations to the case where the variables become operators, $(X,Y,Z)$ has been considered in ref.~\cite{Floratos:2012ys}, specifically,  the model discussed in ref.~\cite{Axenides:2010zr}, where the operators are described by square matrices, that are expanded in the generators of a given Lie algebra. As is usual in quantum mechanics, a prescription for the expressions that involve products of non-commuting variables is mandatory. The Weyl ordering~\cite{Balazs:1984uq} is adopted, so that  the product $XY$ of two operators, $X$ and $Y$  is replaced by its symmetric expression $XY\to \frac{1}{2}\left(XY + YX\right)$.\footnote{It may be checked that another ordering prescription does not change, qualitatively, the results presented here. A more complete discussion will be presented elsewhere.} 
The  $X,Y,Z$ are now expanded in the generators, $T^a$, of a Lie algebra, $\mathcal{G}$, where $a=1,2,\ldots,\mathrm{dim}_{\mathrm{Ad(\mathcal{G})}}$, thus $X\equiv x^a T^a, Y\equiv y^a T^a ~$and$~ Z\equiv z^a T^a$. 
A Lie algebra is defined by its structure constants, $f^{abc}$, that enter in the commutation relations, $[T^a,T^b]=\mathrm{i}f^{abc}T^c$ and the fact that this algebra is compact implies that $\mathrm{Tr}( T^a T^b)= \kappa\delta^{ab}$ with $\kappa\neq 0$ and $\mathrm{Tr}$ stands for the trace on the algebra. 

Here we come upon the following subtlety: The Lie algebra we are considering here, {\em must}  contain a $\mathfrak{u}(1)$ factor. The reason is that, otherwise, the equations are inconsistent, since the trace of the left hand side vanishes, but the trace of the right hand side does not: the trace of the non-linear combinations is non-zero. 

Taking into account  such requirements, the quantum counterparts to eqs.~(\ref{CLS}) take the following, intriguing, form
\begin{equation}
\begin{array}{ccl}
\dot{x}^a&=&\displaystyle{\sigma (y^a - x^a)}\\
\dot{y}^a&=&\displaystyle{-y^a + r x^a -{\sf d}^{abc}x^b z^c}\\
\dot{z}^a&=&\displaystyle{-bz^a + {\sf d}^{abc}x^b y^c}
\end{array}
\label{MLS}
\end{equation}
These equations, herald the appearance of  the invariant symmetric tensor
\begin{equation}
\label{anomaly_tensor}
{\sf d}^{abc}\equiv\frac{1}{2\kappa}\mathrm{Tr}\left[\left\{T^a,T^b\right\}T^c\right]
\end{equation}
of the Lie algebra. This tensor appears in gauge theories, since the gauge fields belong to the adjoint representation of the group and plays an important role in the classification of gauge anomalies~\cite{Georgi:1972fk}. Groups, for which this tensor vanishes identically are called ``anomaly--safe'' and in the present context, this means that eqs.~(\ref{MLS}) are {\em linear}, thus do not give rise to chaotic fluctuations. Only quantum fluctuations can appear. Groups, for which this tensor does not identically vanish, on the other hand, lead to non--linear equations, thus can describe both chaotic {\em and} quantum fluctuations. 


For the case of abelian Lie groups, the structure constants vanish. In that case, eqs.~(\ref{MLS}) is equivalent to eqs.~(\ref{CLS}) up to a rescaling of all the variables proportional to the single non-vanishing element of ${\sf d}$. This means that they share the same route to chaos, i.e. belong to the same universality class. 


When the ${\sf d}$ tensor no longer vanishes, chaotic fluctuations and quantum effects mix in a non-trivial way. The simplest case is the 
$\mathfrak{u}(2)=(\mathfrak{u}(1)\times \mathfrak{su}(2))/\mathbb{Z}_2$ algebra, where all the components of ${\sf d}^{abc}$ are zero, except ${\sf d}^{0cc}=1/2$ for $c\in\{0,1,2,3\}$ with the corresponding circular permutations. This can be made particularly clear by writing the corresponding Lorenz system of equations as follows:

\begin{equation}
\label{SU2U1}
\begin{array}{l}
\mathfrak{u}(1)
\left\{
\begin{array}{l}
\dot{x}^0=\sigma(y^0-x^0)\\
\dot{y}^0=-y^0+r x^0 -x^0z^0-2 x^b z^b\\
\dot{z}^0=-b z^0 + x^0 y^0 + 2x^b y^b
\end{array}
\right.
\\
\\
\mathfrak{su}(2)
\left\{
\begin{array}{l}
\dot{x}^a=\sigma(y^a-x^a)\\
\dot{y}^a = -y^a+r x^a-x^0 z^a-x^a z^0\\
\dot{z}^a = - bz^a + x^a y^0 + x^0 y^a
\end{array}
\right.
\end{array}
\end{equation}

The first set of the three equations highlights the fact that $\mathfrak{u}(1)$ is responsible for the chaotic fluctuations and the second set shows that the $\mathfrak{su}(2)$ components satisfy linear equations, with the $\mathfrak{u}(1)$ variables acting as sources. 

The two sets of equations were integrated using an eighth-order Runge-Kutta scheme with a fixed time stepping scheme. The group invariants, $\mathrm{Tr}(X)$, $\mathrm{Tr}(Y)$ and $\mathrm{Tr}(Z)$, plotted in Fig.~\ref{Fig1}, define a subspace where a structure similar to the Lorenz ``butterfly'' appears, ``decorated'' by the $\mathfrak{su}(2)$ terms that make it ``knotted''. This last property is, in fact, expected \cite{Ghrist:1998uq}, given that great circles on the $\mathfrak{su}(2)$ manifold, the 3--sphere, do have non-zero linking number, the well known Hopf invariant~\cite{Niemi:2014kx}. While knots are stable in the {\em non-chaotic} phase, since they describe periodic orbits, it will be interesting to investigate in more detail  the import of the matrix structure of the equations  on their properties. 

\begin{figure}[htb]
\begin{center}\resizebox{0.4\columnwidth}{!}{\includegraphics{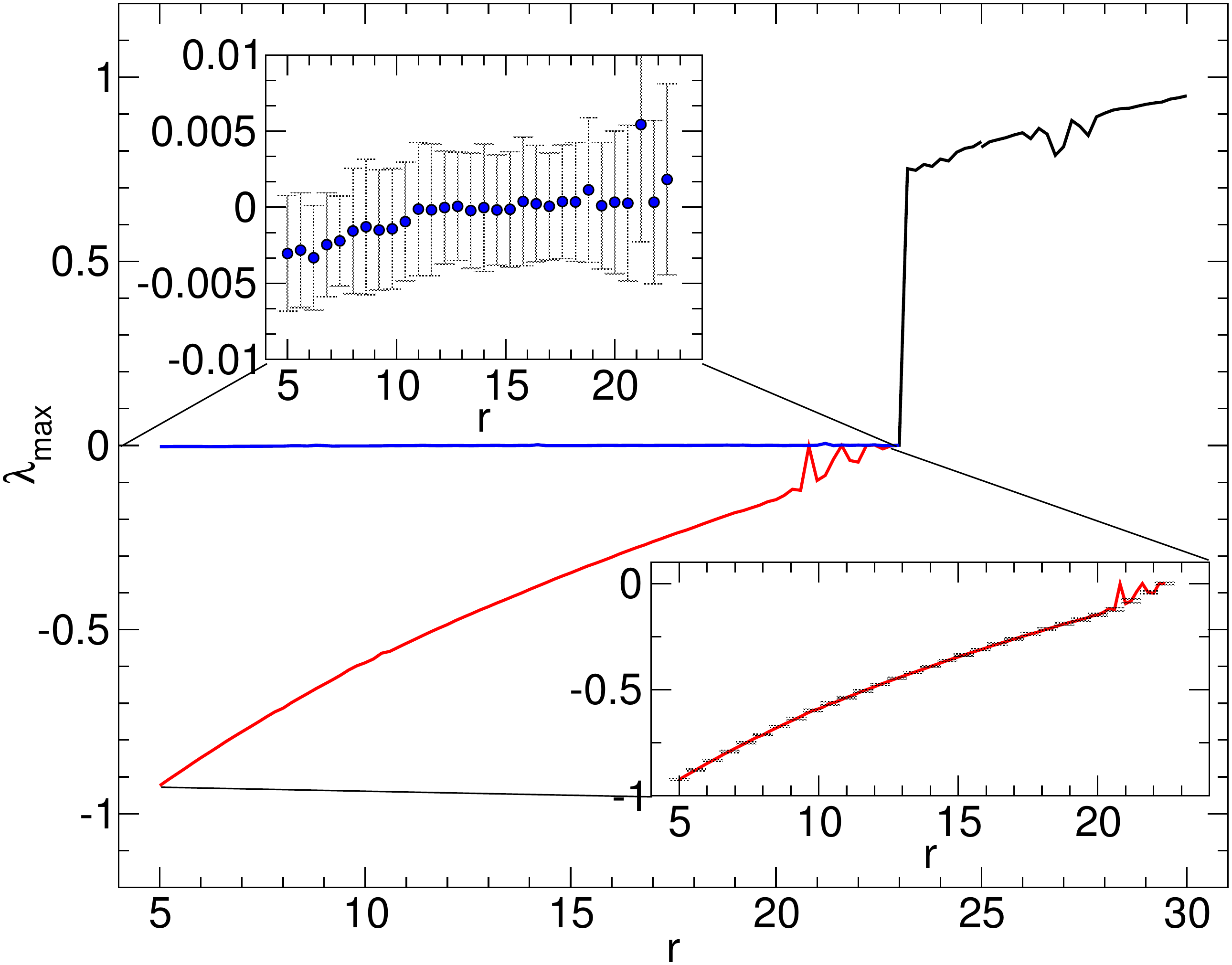}}\end{center}
\caption{(Color online) Phase diagram of the largest Lyapunov spectra in the $\mathfrak{u}(2)$ matrix Lorenz system. Details of selected parts of the diagram are inserted to exhibit the 1$\sigma$ error bars. \label{Fig2}} 
\end{figure}

A striking feature of the largest Lyapunov exponent, as function of $r$, is its bimodal distribution, along the two group factors, 
$\mathfrak{u}(1)$ and $\mathfrak{su}(2)$, in the non-chaotic phase, $r<r_\mathrm{crit}$, whereas, in the chaotic phase, $r>r_\mathrm{crit}$, the distribution 
becomes unimodal, giving the same values for each group factor. While a mathematical proof for this result is not available, we stress that it provides a consistency check for the reliability of our numerical analysis, since the $\mathfrak{su}(2)$ factor cannot give rise to chaos by itself. 
The transition to chaos, at $r=r_\mathrm{crit}$, appears at the same value as for the classical Lorenz system, only much more abrupt in the matrix case. 

\begin{figure}[htb]
\begin{center}\resizebox{0.4\columnwidth}{!}{\includegraphics{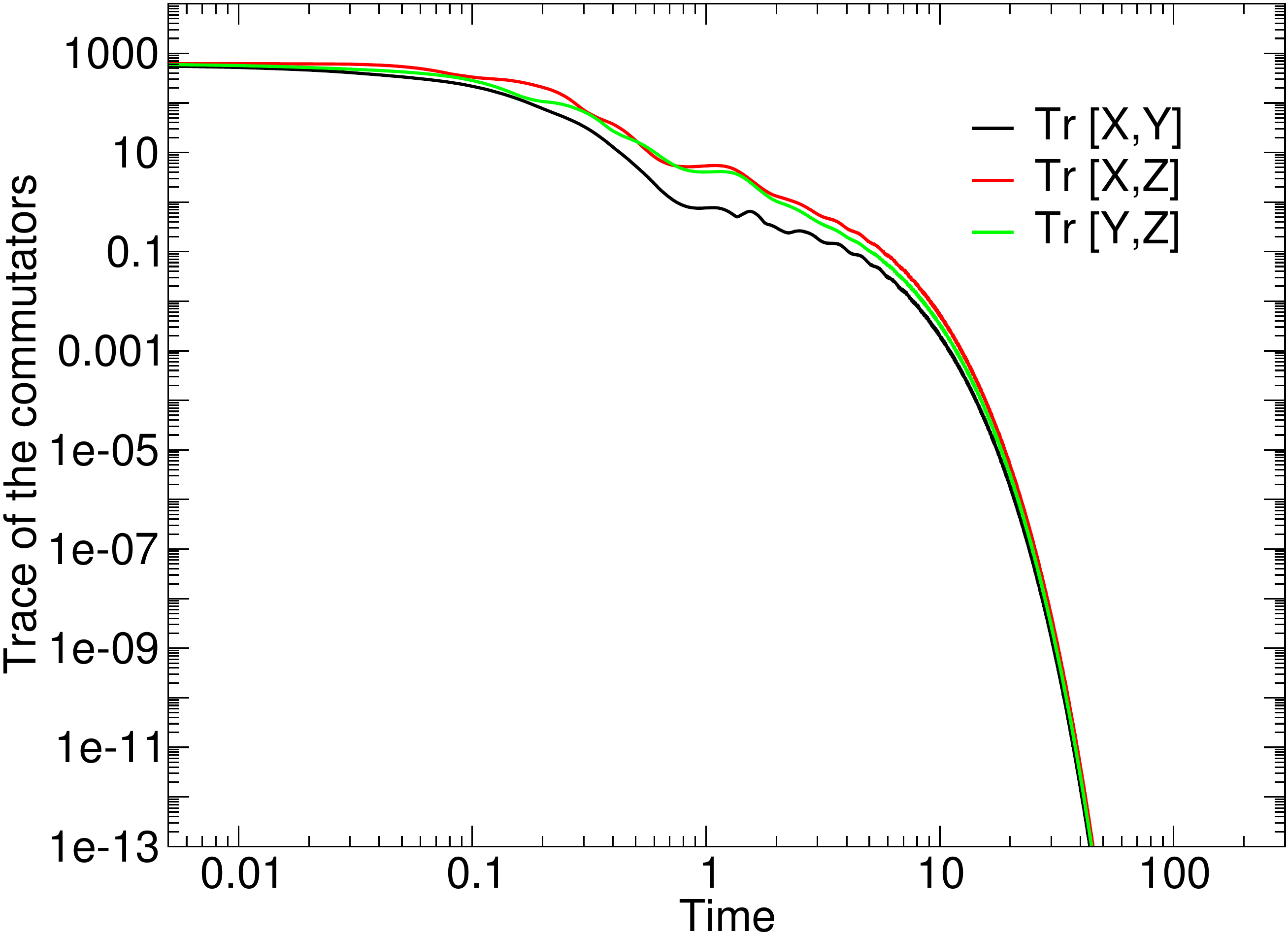}}\end{center}
\caption{(Color online.) Plot of the average of the three commutators ${\mathrm{Tr}}([X,Y]),~{\mathrm{Tr}}([Y,Z]),~{\mathrm{Tr}}([X,Z])$ taken over random initial conditions for $r=15$. \label{LorMatplot}}
\end{figure}

Another way to characterize the quantum fluctuations is by studying the time evolution of the three commutators, ${\mathrm{Tr}}([X,Y])$, ${\mathrm{Tr}}([Y,Z])$ and ${\mathrm{Tr}}([X,Z])$. If all vanish, this means that all three matrices belong to the Cartan subalgebra. Preliminary results show that this does, in fact, occur~\cite{Axenides:2010zr} and our numerical results seem to confirm it.  What is noteworthy in Fig.~\ref{LorMatplot} is that the time evolution of the average of the three commutators taken over random initial conditions, collapses along the same curve.
We shall now eliminate the possibility that the Cartan subalgebra so obtained is proportional only to the identity. This is achieved by monitoring the invariants in the $\mathfrak{su}(2)$ subspace, namely $\mathrm{Tr}X^2-{(\mathrm{Tr}X)}^2={x_1}^2+{x_2}^2+{x_3}^2$ and so on. Performing runs for different values of $r$, at fixed $b$ and $\sigma$, we find that, in the non-chaotic phase, $r<r_\mathrm{crit}$, these quantities seem to converge to a single point, not the origin, whereas in the chaotic phase, $r>r_\mathrm{crit}$, they appear to describe a fuzzy region, of finite volume in phase space. That it does not collapse to the origin indicates the persistence of the quantum fluctuations in both phases. 

A natural generalization of our results is towards a quantum Nambu description of spin systems, where it has been demonstrated that the Nambu mechanics leads to novel identities for extended Lorenz system with dissipation that are not obvious in an Hamiltonian approach \cite{Blender:2013kx}. Moreover this generalization may also provide insights into  the origin of  dissipation in such magnetic systems that has become a subject of topical research~\cite{garanin1997fokker} and lead to new relations that are hard to guess from the Hamiltonian viewpoint.

\ack
JT acknowledges financial support from a joint R\'region Centre--CEA doctoral fellowship. SN would like to thank M Axenides and E G Floratos for stimulating conversations.

\bibliographystyle{iopart-num}
\bibliography{ICM2lorenzQM}
\end{document}